\documentclass[12pt]{article}

\usepackage{graphics,cite,amssymb,epsfig,float,psfrag}
\usepackage[usenames,dvips]{color}
\usepackage{rotating}

\newcommand{\gsim}{\raisebox{-0.13cm}{~\shortstack{$>$ \\[-0.07cm] $\sim$}}~}

\newcommand{\beq}{\begin{eqnarray}} 
\newcommand{\eeq}{\end{eqnarray}} 

\begin{document}
\vspace{1cm}
\begin{center}
{\large\bf Addendum to: Predictions for Higgs production at the Tevatron and the associated
uncertainties}

\vspace{.5cm}

{\large Julien Baglio$^{1}$ and Abdelhak Djouadi$^{1,2}$} 

\vspace*{5mm}

{\small $^1$ Laboratoire de Physique Th\'eorique, Universit\'e Paris-Sud XI et CNRS,
F-91405 Orsay, France.\\
$^2$ Theory Unit, CERN, 1211  Gen\`eve 23, Switzerland.}
\end{center}

\vspace{1cm}

\begin{abstract} In a recent  paper, we updated the theoretical predictions for
the production cross sections of the Standard Model Higgs boson at the 
Tevatron and estimated the various  uncertainties affecting these predictions.
We found that there is a large theoretical uncertainty, of order $40\%$, on the
cross section for the main production  channel, gluon-gluon fusion into a Higgs
boson. Since then, a note from the Higgs working groups of the CDF and D0
collaborations  criticizing  our modeling of the $gg\to H$ cross section has
appeared. In this addendum, we  answer to this criticism point by point and,
in   particular, perform an analysis of $\sigma(gg\to H)$ for a central  value
of the renormalization and factorization scales $\mu_0=\frac12 M_H$ for which
higher order corrections beyond next-to-next-to-leading order (that we
discarded  in our previous analysis) are implicitly included.  Our results 
show that the new Tevatron exclusion bound  on the Higgs boson mass,
$M_{H}\!=\!158$--175 GeV at the 95\% confidence level, is still largely
debatable.

\end{abstract}

\section*{Introduction}

In an earlier paper \cite{Hpaper},  we updated the theoretical predictions for
the production cross sections of the Standard Model Higgs boson at the 
Tevatron collider, focusing on the two main search channels, the gluon--gluon
fusion mechanism $gg\to H$ and the Higgs--strahlung processes $q \bar q \to VH$
with $V=W/Z$, including all relevant higher order perturbative QCD 
\cite{ggH-QCD} and electroweak corrections \cite{ggH-EW,ggH-radja}.  We then
estimated the various theoretical uncertainties affecting these predictions:
the scale uncertainties which are viewed as a measure of the unknown higher
order effects, the uncertainties from the parton distribution functions (PDFs)
and the related errors on the strong coupling constant $\alpha_s$, as well as
the uncertainties due to the use of an effective field theory (EFT) approach in
the determination of the radiative corrections in the $gg\to H$ process at
next-to-next-to-leading order (NNLO). We found that while the cross sections
are well under control in the Higgs--strahlung processes, the uncertainty being
less that $\approx 10\%$, the theoretical uncertainties are rather large in the
case of the gluon--gluon fusion channel, possibly shifting the central values
of the NNLO cross sections by more than $\approx 40\%$. These uncertainties are
thus significantly larger than the $\approx 10\%$ error assumed by the CDF and
D0 experiments in their earlier analysis that has excluded the Higgs mass range
$M_{H}\!=\!162$--166 GeV at the 95\% confidence level (CL) \cite{Tevatron0}. As
$gg\to H$ is by far the dominant  production channel in this mass range, we
concluded that these exclusion limits should be reconsidered in the light of
these large theoretical uncertainties.

After our paper appeared on the archives, some criticisms have been made by the
members  of the Tevatron New Physics and Higgs working group (TevNPHWG) of the
CDF and D0 collaborations \cite{TeVHWG} concerning the theoretical modeling of
the $gg\to H$  production cross section that we proposed. This criticism
appeared on the web in May 2010, but we got aware of it only during ICHEP, i.e.
end of July 2010, where, incidentally, the new combined analysis of CDF and D0
for the Higgs search at the Tevatron was  released \cite{Tev1}. In this
addendum, we respond to this criticism point by point and, in particular,
perform a new analysis of the $gg\to H$ cross section at NNLO for a central 
value of the renormalization and factorization scales $\mu_0=\frac12 M_H$, for
which higher order corrections beyond NNLO (that we discarded with some
justification in our previous analysis) are implicitly included. We take the
opportunity to also comment on the new CDF/D0 results with which the excluded
Higgs mass range was extended to $M_H=$158--175 GeV at the 95\%
CL\footnote{Some of the points that we discuss here have also been presented by
one of us (JB) in the Higgs Hunting workshop in Orsay which followed ICHEP
\cite{HHunting}.}. 

\subsection*{1. The normalization of the $\mathbf{gg\to H}$ cross section} 

One of the points put forward in our paper is to suggest to consider the $gg\to
H$ production cross section up to NNLO \cite{ggH-QCD, ggH-EW,ggH-radja}, 
$\sigma^{\rm NNLO}_{\rm gg\to H}$, and not to include the soft--gluon
resumation contributions \cite{ggH-resum}. The  main reason is that,
ultimately, the observable that is experimentally used is the cross section
$\sigma^{\rm cuts}_{\rm gg\to H}$ in which selection cuts have been applied and
the theoretical prediction for $\sigma^{\rm cuts}_{\rm gg\to H}$ is available
only to NNLO \cite{FEHIP-HNNLO}.  This argument has been criticized by the
TevNPHWG for the reason that we are potentially missing some important higher
order contributions to the cross section. It turns out, however, that our point
is strengthened in the light of the new CDF/D0 combined analysis \cite{Tev1}.
Indeed, in this analysis,  the $gg \to H$  cross section has been broken into
the three pieces which yield different final state signal topologies for the
main decay $H \to WW^{(*0} \to \ell \ell \nu \nu$, namely $\ell \ell \nu
\nu$+0\,jet,  $\ell\ell \nu \nu$+1\,jet and $\ell \ell \nu \nu$+2\,jets or
more: 
\beq
\sigma^{\rm NNLO}_{\rm gg\to H}= \sigma^{\rm 0jet}_{\rm gg\to H}+
\sigma^{\rm 1jet}_{\rm gg\to H}+
\sigma^{\rm \geq2jets}_{\rm gg\to H}
\eeq
These channels have been analyzed separately and these individual components,  
with $\sigma^{\rm 0jet}_{\rm gg\to H}$ evaluated at NNLO, $\sigma^{\rm
1jet}_{\rm gg\to H}$ evaluated at NLO and  $\sigma^{\rm \geq2jets}_{\rm gg\to
H}$ evaluated at LO, represent respectively  $\approx 60\%$, $\approx 30\%$
and  $\approx 10\%$ of the  total $gg\to H$ cross section at NNLO. Since these 
three pieces add up to $\sigma^{\rm NNLO}_{\rm gg\to H}$, we do not find
appropriate to have a different normalization for the jet cross sections and
for the total sum and, thus, to include soft--gluon resumation in the latter
while it is not  taken  into account in the former.

Nevertheless, we are ready to admit that we may have underestimated the total
production cross section, as with the central value of the  renormalization and
factorization  scales $\mu_R=\mu_F=\mu_0=M_H$ that we have adopted for
evaluating $\sigma^{\rm NNLO}_{\rm gg\to H}$, we are missing the $\gsim 10\%$
increase of the cross section due to higher order contributions and, in
particular, to soft--gluon resumation.  As most criticism on our paper focused
on this particular issue (overlooking many other important points that we put
forward), we present here an analysis of the cross section in which these
higher order effects are implicitly taken into account. 

As pointed out by Anastasiou and collaborators some time ago, see e.g. 
Refs.~\cite{ggH-radja,BabbisHH} (and also Ref.~\cite{Review}), the effects of
soft--gluon resumation at NNLL \cite{ggH-resum} can be accounted for in
$\sigma^{\rm NNLO}_{\rm gg\to H}$ by lowering the central value of the 
renormalization and factorization scales\footnote{ Note that the scale choice
$\mu_0=\frac12 M_H$  in $gg\to H$  does not only mimic the inclusion of the
effect of soft--gluon resumation, but it also improves  the convergence of the
perturbative series and is more appropriate  to describe the kinematics of the
process \cite{BabbisHH}.} from  $\mu_0=M_H$ to $\mu_0=\frac12 M_H$. If the
scale value $\mu_0= \frac12 M_H$ is chosen, the central value of $\sigma^{\rm
NNLO}_{\rm gg\to H}$ increases by more than $10\%$ and there is almost no
difference between  $\sigma^{\rm NNLO}_{\rm gg\to H}(\mu_0= \frac12 M_H)$ and
$\sigma^{\rm NNLL}_{\rm gg\to H}(\mu_0=M_H)$ as calculated for instance by de
Florian and Grazzini  \cite{ggH-FG}. 

\begin{figure}[!h]
\vspace*{-.1cm}
\begin{center}
\hspace*{-1.cm}
\epsfig{file=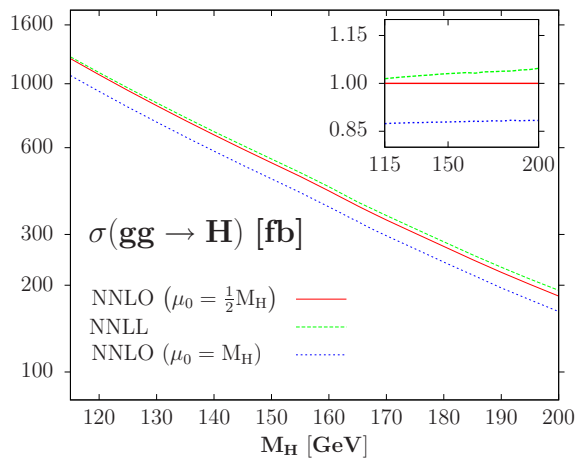,scale=0.7} 
\end{center}
\vspace*{-5mm}
\caption[]{The $gg \to H$ cross section at the Tevatron as  a function of
$M_H$: at NNLO  for central scales at $\mu_{0}=M_H$ and $\mu_0= \frac12 M_{H}$
and at NNLL  for a scale $\mu_0=M_H$. In the insert, shown are
the deviations when one normalizes to $\sigma^{\rm NNLO}_{\rm gg\to
H}(\mu_0=\frac12 M_H)$.}
\vspace*{-1mm}
\label{NF-norm}
\end{figure}

This is explicitly shown in Fig.~\ref{NF-norm} where  $\sigma^{ \rm NNLO}_{\rm
gg\to H}$ with central scales $\mu_0=M_H$ and $\mu_0\!=\!\frac12 M_H$  (that we
calculate following the same lines as the ones discussed in section 2 of our
paper) are compared  to $\sigma^{\rm NNLL}_{\rm gg\to H}$ with $\mu_0\!=\!M_H$
(for which the numbers  are given in  Ref.~\cite{ggH-FG}). For instance, for
$M_H \approx 160$ GeV, while there is a $\simeq 14\%$ difference between
$\sigma^{ \rm NNLO}_{\rm gg\to H}(\mu_0=M_H)$ and $\sigma^{ \rm NNLL}_{\rm
gg\to H}(\mu_0=M_H)$, there is almost no difference between the later and 
$\sigma^{ \rm NNLO}_{\rm gg\to H}(\mu_0=\frac12 M_H)$

As a result of this choice, our normalization for the inclusive $gg\to H$
cross  section is now the same as the ones of  Refs.~\cite{ggH-FG,ggH-radja} 
which were adopted in the combined CDF/D0 analyses. 

\subsection*{2. The scale uncertainty}

The next important issue is the range of variation that one should adopt for
the renormalization and factorization scales, a variation which leads to an
uncertainty band that is supposed  to be a measure of the unknown (not yet
calculated) higher order contributions to the cross section. In our paper, we
have advocated the fact that since the NLO and NNLO QCD corrections in the 
$gg\to H$ process  were so large, it is wiser to extend the range of scale
variation  from what is usually assumed. From the requirement that the scale
variation of the LO or NLO cross sections around the central scale $\mu_0$
catch the central value of $\sigma^{\rm NNLO}_{\rm gg\to H}$, we arrived at the
minimal choice, $\frac13 \mu_0 \leq \mu_R, \mu_F \leq 3 \mu_0$ for
$\mu_{0}=M_{H}$. 

In addition, we proposed that the scales $\mu_R$ and $\mu_F$ are varied 
independently and with no restriction such as $\frac13 \leq \mu_R/\mu_F \leq 3$
for instance.  In fact, this was only a general statement and this requirement
had absolutely no impact on our analysis as the minimal and maximal values of
$\sigma^{\rm NNLO}_{\rm gg\to H}$ due to scale variation were obtained for
equal  $\mu_R$ and $\mu_F$ values: for a central scale $\mu_0=M_H$, one had
$\sigma_{\rm min}^{\rm NNLO}$ for $\mu_R=\mu_F=3 \mu_0$ and $\sigma_{\rm
max}^{\rm NNLO}$ for $\mu_R=\mu_F=\frac13 \mu_0$.

\begin{figure}[!h]
\begin{center}
\hspace*{-1.cm}
\epsfig{file=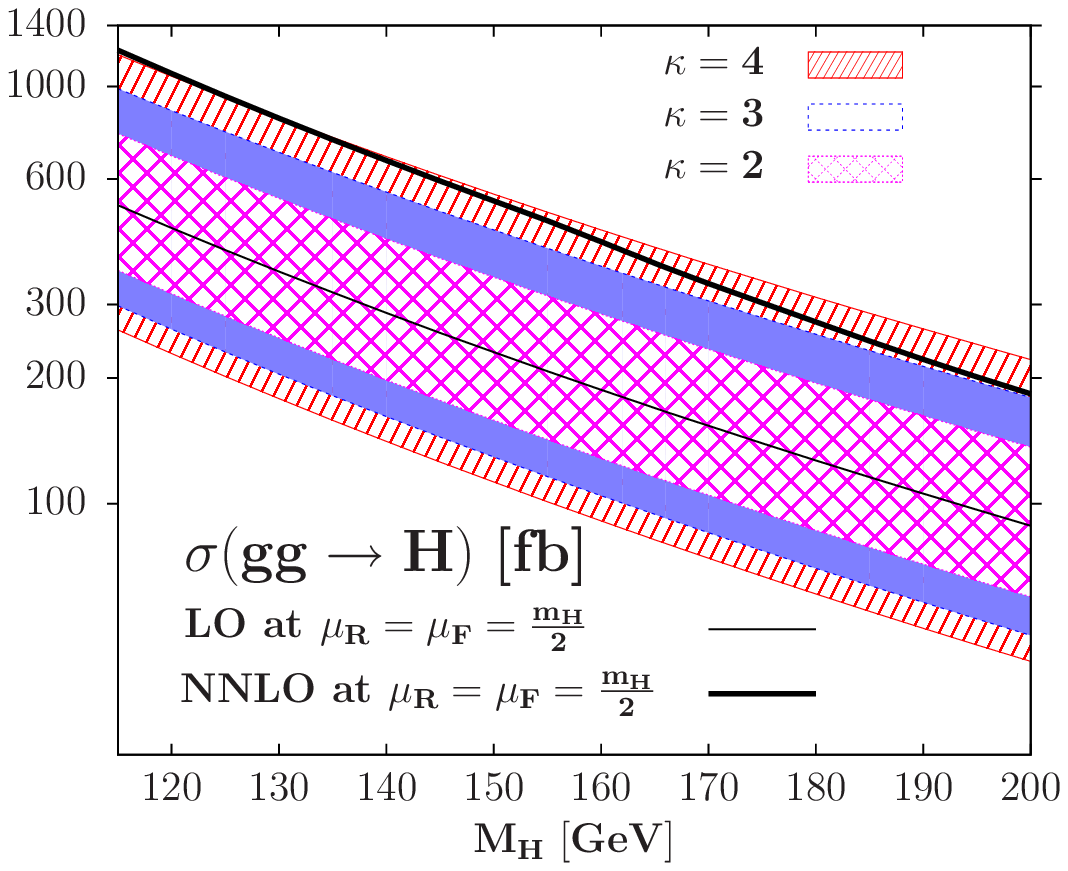,scale=0.7}\hspace*{5mm}
\epsfig{file=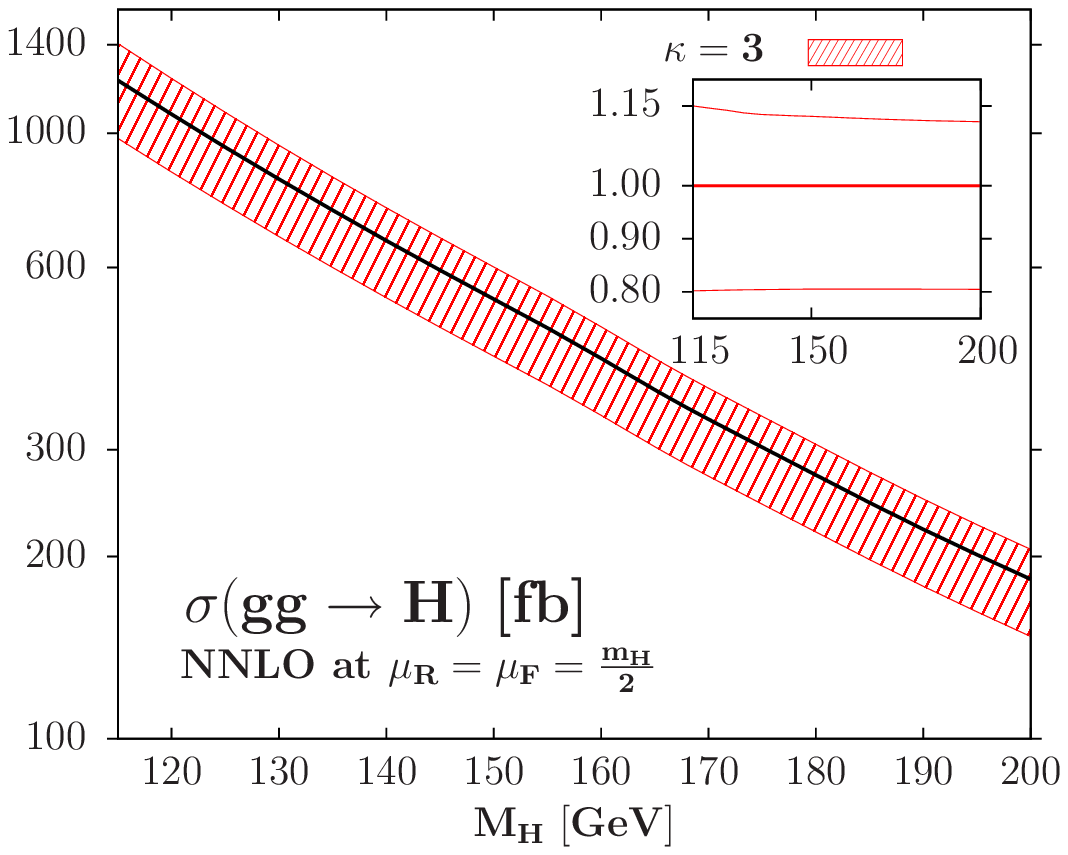,scale=0.7} 
\end{center}
\vspace*{-5mm}
\caption[]{  Left: the scale variation of $\sigma^{\rm LO}_{gg\to H}$  as a
function of $M_H$ in the domain $ \mu_0/\kappa \leq \mu_R=\mu_F \leq \kappa
\mu_0$  for $\mu_0= \frac12 M_H$ with $\kappa=2,3$ and $4$ compared to
$\sigma^{\rm NNLO}_{gg\to H} (\mu_R=\mu_F=\frac12 M_H)$. Right: the uncertainty
band of  $\sigma^{\rm NNLO}_{gg\to H}$ as  a function of $M_H$ for a scale
variation   $\mu_0/\kappa \leq  \mu_R= \mu_F \leq \kappa \mu_0$ with 
$\mu_0=\frac12 M_H$ and $\kappa=3$. In the inserts shown are the relative 
deviations.}
\vspace*{-3mm}
\label{NF-scale}
\end{figure}

Adopting the central scale choice $\mu_0=\frac12 M_H$, for the scale variation
of the  leading--order $gg\to H$ cross section to catch the central value of
$\sigma^{\rm NNLO}_{\rm gg\to H} (\mu_0)$, as shown in the left-hand side of
Fig.~\ref{NF-scale},  we again need to consider the domain 
\beq
\frac13 \mu_0 \leq \mu_R = \mu_F \leq 3 \mu_0 \ , \ \mu_0= \frac12 M_H
\eeq
for the scale variation. Notice that now, we choose for simplicity to equate
$\mu_R$ and $\mu_F$ so that there is no more discussion about the possibility
of generating artificially large logarithms if we take two widely different
$\mu_R/\mu_F$ scales.

Adopting this domain for $\mu_F=\mu_R$, we obtain the result shown in the
right--hand side of  Fig.~\ref{NF-scale}  for the scale variation of the NNLO
cross section around the central scale $\mu_0=\frac12 M_H$.  Averaged over the
entire Higgs mass range, the final scale uncertainty is about $\simeq +15\%,
-20\%$ which, compared with our  previous result for the scale variation of
$\sigma^{\rm NNLO}_{\rm gg\to H}$ with $\mu_0=M_H$ is the same for the minimal
value but smaller for the maximal value. Note that if we had chosen  the usual
domain $\frac12 \mu_0  \leq \mu_R = \mu_F \leq 2 \mu_0$, the scale variation
would have been of about $\approx + 10\%, -12\%$ for $M_H \approx 160$ GeV. 

\vspace{2mm}

It is important to notice that if the NNLO $gg\to H$ cross section, evaluated  
at $\mu_0=M_H$,  is broken into the three pieces with 0,1 and 2 jets, and one
applies a scale variation for the individual pieces in the range $\frac12 \mu_0
\leq \mu_R, \mu_F \leq 2 \mu_0$, one obtains with selection cuts similar to
those adopted by the CDF/D0 collaborations \cite{ggH-ADGSW}: 
\beq
\left.  
\Delta \sigma/\sigma 
\right|_{\rm scale} =  
60\% \cdot \left({^{+5\%}_{-9\%}} \right) 
+29\% \cdot \left({^{+24\%}_{-23\%}} \right) 
+11\% \cdot \left({^{+91\%}_{-44\%}} \right) = \left({^{+20.0\%}_{-16.9\%}} 
\right)
\label{jetscale}
\eeq
Averaged over the various final states with their corresponding weights, an
error on the ``inclusive"  cross section which is about $+20\%,-17\%$ is
derived\footnote{The error might be reduced when including higher--order
corrections in the 1\,jet and 2\,jet cross sections.}. This is very close to
the result obtained in the CDF/D0 analysis  \cite{Tev1} which quotes a scale
uncertainty of $\approx \pm 17.5\%$ on the total cross section, when the
weighted uncertainties for the various jet cross sections are added.  Thus, our
supposedly ``conservative" choice  $\frac13 \mu_0 \leq \mu_R=\mu_F \leq 3
\mu_0$ for the scale variation of the total inclusive cross section
$\sigma^{\rm NNLO}_{\rm gg\to H}$,  leads to  a scale uncertainty that is very
close to that obtained when one adds the scale uncertainties of the various jet
cross sections for a variation around the more ``consensual" range $\frac12
\mu_0 \leq  \mu_R, \mu_F \leq 2 \mu_0$. 

We also note that when breaking $\sigma^{\rm NNLO}_{\rm gg\to H}$ into jet
cross sections, an additional error due to the acceptance of jets is
introduced; the CDF and D0 collaborations, after weighting, have estimated it
to be $\pm 7.5\%$. We do not know if this weighted acceptance error should be
considered as a theoretical or an experimental uncertainty. But this error,
combined with the weighted uncertainty for scale variation, will certainly
increase the total scale error in the CDF/D0  analysis, possibly (and depending
on how the errors should be added) to the level where it almost reaches or even
exceeds our own supposedly ``conservative" estimate. 

\subsection*{3. PDF and $\alpha_{s}$ uncertainties}

Another issue is the uncertainties due to the parameterization of the PDFs and
the corresponding ones from the value of the strong coupling constant
$\alpha_s$. In their updated analysis \cite{Tev1}, the CDF and D0
collaborations are now including the uncertainties generated by the
experimental error in the value of $\alpha_s$ and considering the
PDF+$\Delta^{\rm exp}\alpha_{s}$ uncertainty, but there is still a little way
to go as the problem of the theoretical error  on $\alpha_s$ is still pending. 

\begin{figure}[!h]
\begin{center}
\epsfig{file=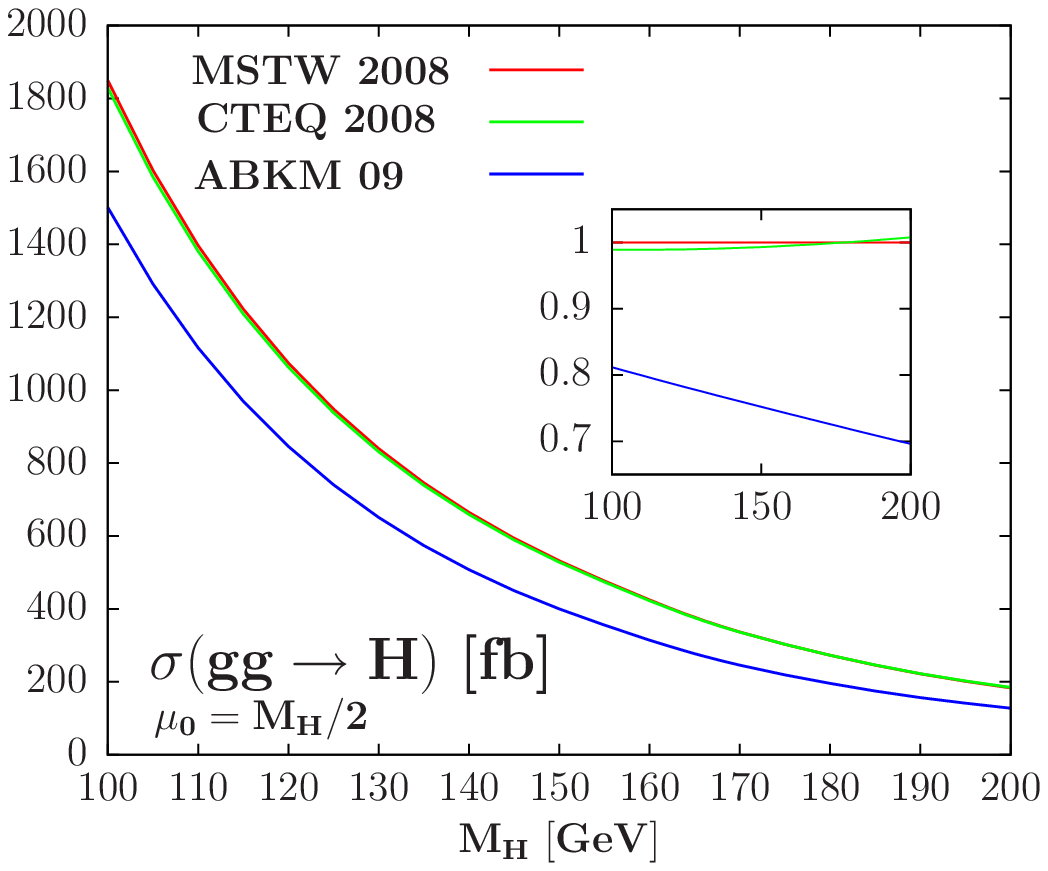,scale=0.7}\hspace*{5mm} 
\epsfig{file=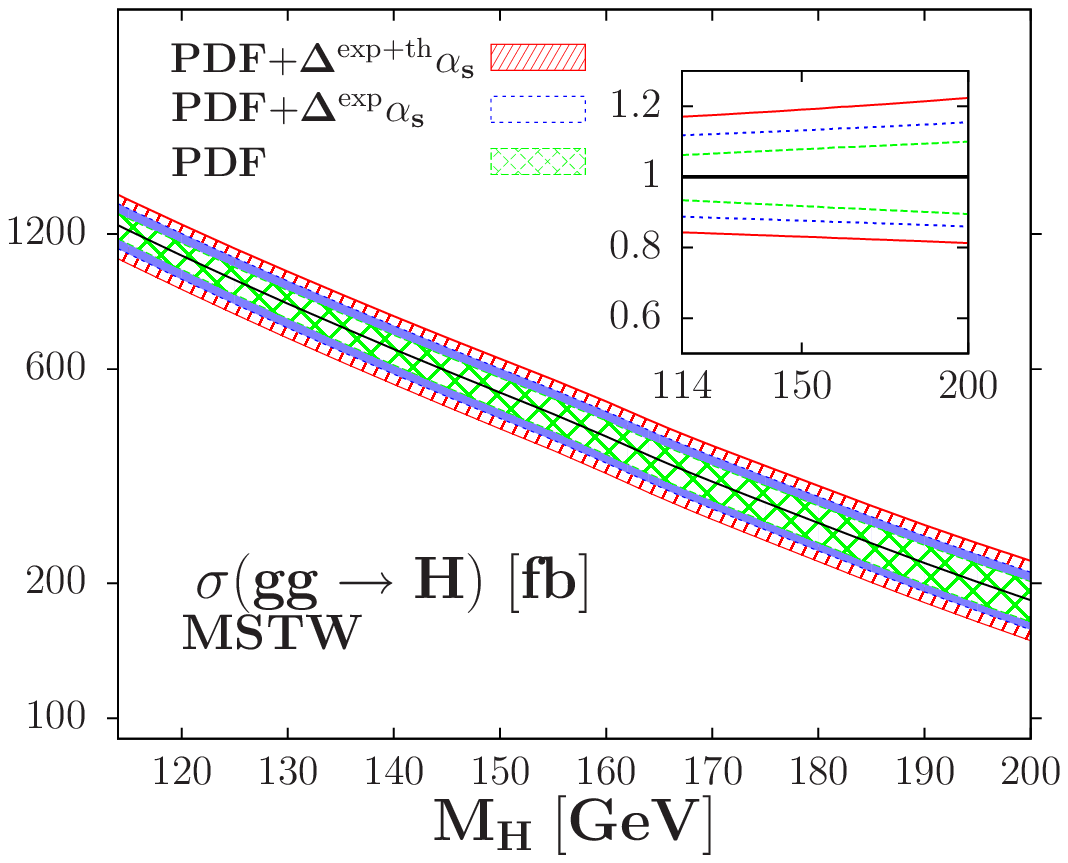,scale=0.7} 
\end{center}
\vspace*{-5mm}
\caption[]{Left: the $gg\to H$ cross section at NNLO for $\mu_{0}=\frac12 M_{H}$
as a function of $M_H$ when the  MSTW, CTEQ and ABKM parameterizations are used.
Left: the 90\%CL PDF, PDF+$\Delta^{\rm exp}\alpha_{s}$ and PDF+$ \Delta^{\rm 
exp+th}\alpha_{s}$ uncertainties on $\sigma^{\rm NNLO}_{gg\to H}$ in the MSTW 
parametrisation. In the inserts, shown  in \% are the deviations with
respect to the central MSTW value.}
\vspace*{1mm}
\label{NF-PDF}
\end{figure}

For the new analysis that we present here for $\sigma^{\rm NNLO}_{\rm gg\to
H}$  with a central scale $\mu_0=\frac12 M_H$,  we have only slightly changed
our previous recipe for calculating the errors due to PDFs and $\alpha_s$: we
still use the grids provided by the MSTW collaboration \cite{PDF-MSTW} for
PDF+$\Delta^{\rm exp} \alpha_{s}$, take the 90\%CL  result and add in
quadrature the impact of the theoretical error $\Delta^{\rm th} \alpha_{s}$
using again  the sets provided by the MSTW collaboration. However, contrary to
the case $\mu_0=M_H$ where the value $\Delta^{\rm th} \alpha_{s}=0.003$ at NLO
($\Delta^{\rm th} \alpha_{s}=0.002$ at NNLO) as suggested by  MSTW
\cite{PDF-MSTW} was sufficient to achieve a partial overlap of  the MSTW and
ABKM predictions (which, together with the CTEQ prediction,  are given in the
left--hand side of Fig.~\ref{NF-PDF}) when including their respective error
bands,  we need in the case $\mu_0=\frac12 M_H$ an uncertainty of $\Delta^{\rm
th} \alpha_{s}=0.004$, to make such that the MSTW and ABKM predictions, which
differ by more than 25\% in this case, become consistent. 

Adopting this value for the $\alpha_s$ theoretical uncertainty,  which is
approximately  the difference between the MSTW and ABKM central  $\alpha_s$
values, the results for  $\sigma^{\rm NNLO}_{gg\to H}$ using only the MSTW 
parametrisation are displayed in the right--hand side of Fig.~\ref{NF-PDF}.
Shown  are  the 90\% confidence level  PDF, PDF+$\Delta^{\rm exp}\alpha_{s}$
and PDF+$\Delta^{\rm exp+th}\alpha_{s}$ uncertainties, with the 
PDF+$\Delta^{\rm exp}\alpha_{s}$ and PDF+$\Delta^{\rm th}\alpha_{s}$ combined
in quadrature.  We thus obtain a PDF+$\Delta^{\rm exp+th}\alpha_{s}$ total
uncertainty of $\pm 15\%$ to  $20\%$  on the central cross section depending on
the $M_H$ value. This is larger than the 12.5\% error which has been assumed in
the most recent CDF/D0 combined analysis \cite{Tev1} (and even larger than the
$\approx \pm 8\%$ assumed in the earlier analysis \cite{Tevatron0}).  We
believe that if the  effect of the theoretical error on $\alpha_s$ is taken
into account in the Tevatron analysis of  $\sigma(gg\to H)$, we will arrive at
a much closer agreement. 

We would like to insist on the fact  that this recipe is only one particular
way, and by no means the only one, of parameterizing the PDF  uncertainty.  A
possibly more adequate procedure to evaluate this  theoretical uncertainty
would be to consider the difference between the central values given by various
PDF sets. In the present $gg\to H$ case, while the MSTW and CTEQ
parameterizations give approximately  the same result as shown previously, ABKM
gives a central NNLO cross section that is $\approx 25\%$ smaller  than that
obtained using the  MSTW set\footnote{The   $gg\to H$ cross section is even
smaller  if one uses the new NNLO central PDF sets recently released by the
HERAPDF collaboration  \cite{HERAPDF} rather  than the  ABKM PDF set.}. The PDF
uncertainty, in this case, would be thus $\approx -25\%,+0\%$. 

We also note that there is another recipe  that has been suggested by the
PDF@LHC working group for evaluating PDF  uncertainties for NNLO cross sections
(besides taking the envelope of the predicted values obtained using several PDF
sets) \cite{Thorne}:  take the MSTW PDF+$\Delta^{\rm exp} \alpha_s$ error and
multiply it by a factor of two. In our case, this would lead to an  uncertainty
of $\approx \pm 25\%$ which, for the minimal value, is close to the recipe
discussed just above, and is larger than what we obtain when considering  the
PDF+$\Delta^{\rm exp+th} \alpha_s$ uncertainty given by MSTW.    We thus
believe that our estimate of the PDF+$\alpha_s$ uncertainty that we quote here
is far from being exaggeratedly conservative.

\subsection*{4. Combination of the various uncertainties}

The last issue that remains to be discussed and which, to our opinion is  the
main one, is the way of combining the various sources of theoretical errors. 
Let us first reiterate an important comment: the uncertainties associated to
the PDF parameterisations are theoretical errors and they have been considered
as such since a long time.  Indeed, although the PDF sets use various
experimental data which have intrinsic errors (and which are at the origin of
the misleading ``probabilistic"  interpretation  of the errors given by each
PDF set that are generally quoted),  the main uncertainty is due to the
theoretical  assumptions which go into the different parameterizations. This
uncertainty cannot be easily  quantified within one given parametrisation but
it is reflected in the spread of the central values given by the various PDF
parameterizations that are available. If one defines the PDF uncertainty as the
difference in the cross sections when using the different available PDF sets, 
this uncertainty has no statistical or probabilistic ground. For the scale
uncertainty, the situation is of course clear: it has no statistical ground and
any value of the cross section in the uncertainty band is as respectable as
another\footnote{In  statistical language, both the scale and PDF uncertainties
have a flat prior. A more elaborated discussion on this issue  will appear in a
separate publication \cite{combination}.}.  

As a result, the scale and PDF uncertainties,  cannot be  combined in
quadrature as done, for instance, by the CDF and D0 collaborations.  This is
especially true as in the $gg\to H$ process, a strong correlation between the 
renormalization and factorization scales that are involved (and that we have
equated here for simplicity, $\mu_R\!=\!\mu_F$), the value of $\alpha_s$ and
the $gg$ densities is present. For instance, decreasing (increasing)  the 
scales will increase (decrease) the $gg \to H$ cross section not only because
of the lower (higher)  $\alpha_s (\mu_R^2)$ value that is obtained and which
decreases (increases) the magnitude of the matrix element squared (that is
proportional to $\alpha_s^2$ at leading order  and  the cross section is
minimal/maximal for the highest/lowest $\mu_R=\mu_F$ values), but also because
at the same time,  the $gg$ densities become smaller (larger)  for higher
(smaller) $\mu_F\!=\!\sqrt{Q^2}$ values. See Ref.~\cite{Thorne} for details. 

Thus, not only the scale and PDF uncertainties cannot be added in quadrature, 
they also cannot be added linearly because of the aforementioned correlation. 
We therefore strongly believe that the best and safest procedure to combine
the  scale and PDF+$\alpha_s$  uncertainties is the one proposed in our paper,
that is, to estimate  directly the PDF+$\alpha_s$ uncertainties on the maximum
and minimum cross  sections with respect to the scale variation, $\sigma_0 \pm 
\Delta \sigma^+_\mu$. 

In addition, there is a last theoretical error which should be included, 
related to the use of the EFT approach for the b--quark loop at NNLO QCD 
(together with the parametric and scheme uncertainty on the $b$--quark mass) 
and  for the electroweak radiative corrections, which amount to a few \%. These
uncertainties, discussed in detail in section 3.2 of our paper, are also purely
theoretical uncertainties and should be added linearly to the combined scale 
and PDF+$\alpha_s$ uncertainty (as there is no apparent correlation between
them).  

\begin{figure}[!h]
\vspace*{1mm}
\begin{center}
\epsfig{file=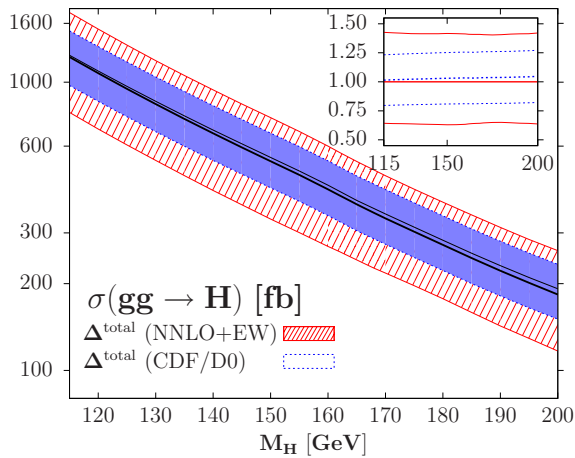,scale=0.7}
\end{center}
\vspace*{-5mm}
\caption[]{The production cross section $\sigma(gg\to H)$ at NNLO for the QCD
and NLO for the electroweak corrections at the Tevatron at a central scale 
$\mu_F=\mu_R= \frac12 M_H$ with the uncertainty band when all theoretical
uncertainties are added using our procedure. It is compared to $\sigma(gg 
\to H)$ at NNLL~\cite{ggH-FG} with the errors quoted by the CDF/D0 
collaboration \cite{Tev1}. In the insert, the relative deviations 
compared to the central value are shown.} 
\label{NF-total}
\vspace*{-2mm}
\end{figure}

Doing so for the $gg\to H$ NNLO cross section with a central scale
$\mu_0=\frac12 M_H$, we obtain the total error shown in Fig.~\ref{NF-total},
that we compare to the $ \approx \pm 22\%$ error assumed in the CDF/D0
analysis.  For $M_H= 160$ GeV for instance, we obtain $\Delta \sigma/\sigma
\approx +41\%, - 37\%$. Compared to our previous result with a central scale
$\mu_0=M_H$ which amounted to  $\Delta \sigma/\sigma \simeq +48\%,-40\%$, this
is approximately the same (only a few percent less) for the lower value of the
cross section and significantly less for its upper value.

Hence, our procedure for the combination does not reduce to a linear sum of all
uncertainties. If we had added linearly all errors, we would have 
had, for the negative part at $M_H=160$ GeV,  a total uncertainty of $\Delta
\sigma/\sigma \approx - 42\%$, compared to  the value $- 37\%$ with our
procedure. On the other hand, one has an error of $\approx - 30\%$, i.e. close
to the total error assumed by CDF/D0 if the scale and PDF+$\alpha_s$
uncertainties were added in quadrature and the EFT approach error linearly 
(the latter being ignored by the CDF/D0 collaborations). 

\subsection*{Summary} 

We have updated our analysis on the theoretical predictions for the Higgs
production cross section in the $gg\to H$ process at the Tevatron, by assuming 
a central scale $\mu_R=\mu_F=\mu_0=\frac12 M_H$ which seems more  appropriate
to describe the process and implicitly accounts for the bulk of the higher
order contributions  beyond NNLO. We have then estimated the theoretical
uncertainties associated to the  prediction: the scale uncertainty, the
uncertainties from the PDF parametrisation and the associated error on
$\alpha_s$, as well as uncertainties due to the use of the EFT approach for the
mixed QCD-electroweak  radiative corrections and the $b$-quark loop
contribution. In Table \ref{NF-tab}, we summarise the results that we have
obtained: the first column shows the central cross section obtained at NNLO
with $\mu_{0} = \frac12 M_{H}$ and the other columns the individual
uncertainties and the total absolute and relative uncertainties when the latter
are combined using our procedure.

While our central value agrees now with the ones given in
Refs.~\cite{ggH-FG,ggH-radja} and adopted by the CDF/D0 collaborations, the
overall theoretical uncertainty that we obtain is approximately twice the error
assumed in the latest Tevatron analysis  to obtain the exclusion band  158 GeV
$\leq M_H \leq$ 175 GeV on the Higgs mass \cite{Tev1}. This is a mere
consequence of the different ways to combine the individual scale and
PDF+$\alpha_s$ uncertainties  and, to a lesser extent, the impact on the
theoretical uncertainty on $\alpha_s$ and the EFT uncertainties which have not
been considered  by the CDF/D0 collaborations. We have provided arguments in
favor of our procedure to combine the scale and PDF uncertainties and we
therefore still believe that the CDF/D0 exclusion limit on the Higgs mass
should be reconsidered.

\begin{table}[!h]{\small%
\let\lbr\{\def\{{\char'173}%
\let\rbr\}\def\}{\char'175}%
\renewcommand{\arraystretch}{1.5}
\vspace*{10mm}
\begin{center}
\begin{tabular}{|c||c||c|ccc|cc||cc|}\hline
$~~M_H~~$ & $\sigma^{\rm NNLO}_{\rm gg \to H}$ [fb]&~~scale~~&PDF& PDF+$
\alpha_s^{\rm exp}$&$\alpha_s^{\rm th}$& EW & b--loop & total & \% total \\ \hline
$100$ & 1849 & $^{+318}_{-371}$ & $^{+102}_{-109}$ & $^{+210}_{-201}$ &
$^{+219}_{-199}$ & $^{+45}_{-45}$ & $^{+42}_{-42}$ & $^{+817}_{-648}$ & $^{+44.2\%}_{-35.0\%}$ \\ \hline
$105$ & 1603 & $^{+262}_{-320}$ & $^{+91}_{-98}$ & $^{+184}_{-176}$ &
$^{+192}_{-174}$ & $^{+41}_{-41}$ & $^{+39}_{-39}$ & $^{+700}_{-565}$ & $^{+43.7\%}_{-35.3\%}$ \\ \hline
$110$ & 1397 & $^{+219}_{-277}$ & $^{+83}_{-89}$ & $^{+163}_{-156}$ &
$^{+170}_{-152}$ & $^{+37}_{-37}$ & $^{+35}_{-35}$ & $^{+602}_{-496}$ & $^{+43.1\%}_{-35.5\%}$ \\ \hline
$115$ & 1222 & $^{+183}_{-242}$ & $^{+75}_{-81}$ & $^{+144}_{-138}$ &
$^{+151}_{-134}$ & $^{+33}_{-33}$ & $^{+32}_{-32}$ & $^{+521}_{-437}$ & $^{+42.6\%}_{-35.7\%}$ \\ \hline
$120$ & 1074 & $^{+156}_{-211}$ & $^{+69}_{-73}$ & $^{+129}_{-123}$ &
$^{+135}_{-119}$ & $^{+30}_{-30}$ & $^{+29}_{-29}$ & $^{+454}_{-386}$ & $^{+42.2\%}_{-36.0\%}$ \\ \hline
$125$ & 948 & $^{+134}_{-186}$ & $^{+63}_{-67}$ & $^{+115}_{-110}$ &
$^{+121}_{-106}$ & $^{+28}_{-28}$ & $^{+24}_{-24}$ & $^{+397}_{-342}$ & $^{+41.9\%}_{-36.1\%}$ \\ \hline
$130$ & 839 & $^{+115}_{-164}$ & $^{+57}_{-61}$ & $^{+104}_{-99}$ &
$^{+108}_{-94}$ & $^{+25}_{-25}$ & $^{+21}_{-21}$ & $^{+349}_{-304}$ & $^{+41.5\%}_{-36.2\%}$ \\ \hline
$135$ & 746 & $^{+100}_{-145}$ & $^{+53}_{-56}$ & $^{+94}_{-89}$ &
$^{+98}_{-84}$ & $^{+23}_{-23}$ & $^{+18}_{-18}$ & $^{+309}_{-272}$ & $^{+41.4\%}_{-36.5\%}$ \\ \hline
$140$ & 665 & $^{+88}_{-129}$ & $^{+48}_{-51}$ & $^{+85}_{-80}$ &
$^{+88}_{-76}$ & $^{+21}_{-21}$ & $^{+16}_{-16}$ & $^{+275}_{-243}$ & $^{+41.4\%}_{-36.6\%}$ \\ \hline
$145$ & 594 & $^{+78}_{-115}$ & $^{+45}_{-47}$ & $^{+77}_{-73}$ &
$^{+80}_{-68}$ & $^{+19}_{-19}$ & $^{+14}_{-14}$ & $^{+246}_{-218}$ & $^{+41.4\%}_{-36.8\%}$ \\ \hline
$150$ & 532 & $^{+69}_{-103}$ & $^{+41}_{-44}$ & $^{+70}_{-66}$ &
$^{+73}_{-61}$ & $^{+17}_{-17}$ & $^{+13}_{-13}$ & $^{+221}_{-197}$ & $^{+41.6\%}_{-37.0\%}$ \\ \hline
$155$ & 477 & $^{+61}_{-92}$ & $^{+38}_{-40}$ & $^{+64}_{-60}$ &
$^{+67}_{-55}$ & $^{+15}_{-15}$ & $^{+10}_{-10}$ & $^{+198}_{-176}$ & $^{+41.5\%}_{-37.0\%}$ \\ \hline
$160$ & 425 & $^{+54}_{-82}$ & $^{+35}_{-37}$ & $^{+58}_{-54}$ &
$^{+60}_{-50}$ & $^{+11}_{-11}$ & $^{+9}_{-9}$ & $^{+175}_{-155}$ & $^{+41.3\%}_{-36.6\%}$ \\ \hline
$162$ & 405 & $^{+51}_{-78}$ & $^{+33}_{-35}$ & $^{+56}_{-52}$ &
$^{+58}_{-48}$ & $^{+9}_{-9}$ & $^{+8}_{-8}$ & $^{+166}_{-146}$ & $^{+40.9\%}_{-36.2\%}$ \\ \hline
$164$ & 386 & $^{+48}_{-75}$ & $^{+32}_{-34}$ & $^{+53}_{-50}$ &
$^{+55}_{-45}$ & $^{+8}_{-8}$ & $^{+8}_{-8}$ & $^{+158}_{-139}$ & $^{+40.9\%}_{-36.0\%}$ \\ \hline
$165$ & 377 & $^{+47}_{-73}$ & $^{+31}_{-33}$ & $^{+52}_{-48}$ &
$^{+54}_{-44}$ & $^{+7}_{-7}$ & $^{+8}_{-8}$ & $^{+154}_{-135}$ & $^{+40.8\%}_{-35.9\%}$ \\ \hline
$166$ & 368 & $^{+46}_{-71}$ & $^{+31}_{-33}$ & $^{+51}_{-47}$ &
$^{+53}_{-44}$ & $^{+6}_{-6}$ & $^{+8}_{-8}$ & $^{+150}_{-132}$ & $^{+40.9\%}_{-35.8\%}$ \\ \hline
$168$ & 352 & $^{+44}_{-68}$ & $^{+30}_{-31}$ & $^{+49}_{-46}$ &
$^{+51}_{-42}$ & $^{+5}_{-5}$ & $^{+8}_{-8}$ & $^{+144}_{-126}$ & $^{+40.9\%}_{-35.7\%}$ \\ \hline
$170$ & 337 & $^{+42}_{-65}$ & $^{+29}_{-30}$ & $^{+47}_{-44}$ &
$^{+49}_{-40}$ & $^{+4}_{-4}$ & $^{+7}_{-7}$ & $^{+137}_{-119}$ & $^{+40.6\%}_{-35.4\%}$ \\ \hline
$175$ & 303 & $^{+37}_{-59}$ & $^{+26}_{-28}$ & $^{+43}_{-40}$ &
$^{+45}_{-36}$ & $^{+2}_{-2}$ & $^{+6}_{-6}$ & $^{+122}_{-106}$ & $^{+40.4\%}_{-35.1\%}$ \\ \hline
$180$ & 273 & $^{+33}_{-53}$ & $^{+24}_{-26}$ & $^{+39}_{-36}$ &
$^{+41}_{-33}$ & $^{+1}_{-1}$ & $^{+6}_{-6}$ & $^{+111}_{-95}$ & $^{+40.6\%}_{-34.9\%}$ \\ \hline
$185$ & 245 & $^{+30}_{-47}$ & $^{+22}_{-24}$ & $^{+36}_{-33}$ &
$^{+38}_{-30}$ & $^{+1}_{-1}$ & $^{+6}_{-6}$ & $^{+101}_{-87}$ & $^{+41.1\%}_{-35.3\%}$ \\ \hline
$190$ & 222 & $^{+27}_{-43}$ & $^{+21}_{-22}$ & $^{+33}_{-30}$ &
$^{+35}_{-27}$ & $^{+2}_{-2}$ & $^{+5}_{-5}$ & $^{+92}_{-79}$ & $^{+41.4\%}_{-35.7\%}$ \\ \hline
$195$ & 201 & $^{+24}_{-39}$ & $^{+19}_{-20}$ & $^{+31}_{-28}$ &
$^{+32}_{-25}$ & $^{+2}_{-2}$ & $^{+3}_{-3}$ & $^{+83}_{-72}$ & $^{+41.4\%}_{-35.8\%}$ \\ \hline
$200$ & 183 & $^{+22}_{-35}$ & $^{+18}_{-19}$ & $^{+28}_{-26}$ &
$^{+30}_{-23}$ & $^{+2}_{-2}$ & $^{+3}_{-3}$ & $^{+77}_{-67}$ & $^{+42.0\%}_{-36.3\%}$ \\ \hline
\end{tabular} 
\end{center} 
\caption{The NNLO total Higgs production cross sections in the $\protect{gg\to
H}$  process at the Tevatron (in fb) for given Higgs mass values (in GeV) at a
central scale $\mu_F=\mu_R=\frac12 M_H$. Shown also are the  corresponding
shifts due to the theoretical uncertainties from the various sources discussed, as well as the total uncertainty when all errors are added using
the  procedure described in the text. } 
\label{NF-tab}
\vspace*{-1mm}
}
\end{table}

\end{document}